\documentclass[12pt]{article}

\usepackage{amssymb}
\usepackage{amsbsy}

\newcommand{\Real}{\mathop{\mathbb R}\nolimits}           
\newcommand{\posReal}{\mathop{\mathbb R}^+\nolimits}      
\newcommand{\ie}{\emph{i.e.}}                             
\newcommand{\eg}{\emph{e.g.}}                             
\newcommand{\cf}{\emph{cf.}}                              
\newcommand{\etc}{\emph{etc.}}                            
\newcommand{\qed}
           {\mbox{\quad\rule[-1.5pt]{.4em}{1.5ex}}}       
\newcommand{\rhs}{\emph{r.h.s.}~}                         
\newcommand{\supp}{\mathop{\mathrm{supp}}\nolimits}       
\newcommand{\Tr}{\mathop{\mathrm{Tr}}\nolimits}           
\newcommand{\Tot}{\mathop{\mathsf{Tot}}\nolimits}         
\newcommand{\diag}{\mathop{\mathrm{diag}}\nolimits}       

\newcommand{\si}{\boldsymbol{L}^{1}}                      
\newcommand{\sii}{\boldsymbol{L}^{2}}                     
\newcommand{\sobii}{\boldsymbol{W}_{2}^{2}}               
\newcommand{\Comp}{\boldsymbol{C}_0^\infty}               
\newcommand{\Diff}{\boldsymbol{C}^\infty}                 
\newcommand{\Diffii}{\boldsymbol{C}^2}                    

\newcommand{\calA}{\mathcal{A}}               
\newcommand{\calB}{\mathcal{B}}               
\newcommand{\calC}{\mathcal{C}}               
\newcommand{\calP}{\mathcal{P}}               

\newcommand{\PF}{\textsc{Proof:\quad}}                    
\newtheorem{claim}{Claim}[section]
\newtheorem{prop}[claim]{Proposition}                     
\newtheorem{thm}[claim]{Theorem}                          
\newtheorem{corol}[claim]{Corollary}                      
\newtheorem{rem}[claim]{Remark}                           

\newcommand{\beq}{\begin{equation}}                       
\newcommand{\eeq}{\end{equation}}                         
\newcommand{\beqn}{\begin{eqnarray}}                      
\newcommand{\eeqn}{\end{eqnarray}}                        

\def\OMIT#1{}                         

\begin{document}


\title{Locally curved quantum layers}
\author{P.~Duclos$^{a,b}$,
        P.~Exner$^{c,d}$,
        and D.~ Krej\v{c}i\v{r}\'{\i}k$^{a,b,c,e}$}
\date{} \maketitle

{\small\em
\begin{description}
\item[$a)$] Centre de Physique Th\'eorique, CNRS,
            13288 Marseille-Luminy
            \vspace{-1.2ex}
\item[$b)$] PHYMAT, Universit\'e de Toulon et du Var,
            83957 La Garde, France
            \vspace{-1.2ex}
\item[$c)$] Nuclear Physics Institute, Academy of Sciences,
            25068 \v{R}e\v{z} near Prague
            \vspace{-1.2ex}
\item[$d)$] Doppler Institute, Czech Technical University,
            B\v{r}ehov{\'a}~7, 11519 Prague
            \vspace{-1.2ex}
\item[$e)$] Faculty of Mathematics and Physics, Charles University,
            V~Hole\v{s}ovi\v{c}k\'ach~2, 18000 Prague, Czech Republic
            \vspace{-1.2ex}
\end{description}
}
\smallskip
\textrm{duclos@univ-tln.fr, exner@ujf.cas.cz, krejcirik@ujf.cas.cz}
\bigskip


\begin{quote}
\noindent {\small We consider a quantum particle constrained to a
curved layer of a constant width built over an infinite smooth
surface. We suppose that the latter is a locally deformed plane
and that the layer has the hard-wall boundary. Under this
assumptions we prove that the particle Hamiltonian possesses
geometrically induced bound states.}
\end{quote}

\setcounter{equation}{0}
\section{Introduction}

Relations between geometry and spectral properties are one of the
trademark topic of mathematical physics, in fact, an abstraction
of various acoustic problems can be found in the roots of this
discipline. For a long time, however, it seemed that such
questions were restricted to the area of classical physics. This
was mostly because geometrical properties of quantum systems were
supposed to result from their dynamics, and as such they were not
accessible to experimenters choice and manipulation. Interesting
results dependent on geometry existed, of course, such as chaotic
behaviour of certain quantum billiards observed first
in~\cite{BGS} and studied in numerous subsequent papers, but they
remained to be mostly academic exercises.

The situation has changed dramatically with the advent of
mesoscopic techniques which allow us to produce tiny structures of
various shapes devised and reproducible in the laboratory and yet
small enough to exhibit quantum effects. Moreover, the physical
nature of these objects makes it possible to describe them by a
simple model in which a free particle (with an effective mass) is
confined to the spatial region of the structure -- see \cite{DE1}
and references therein -- borrowing the terminology one can speak
about quantum waveguides, resonators, \etc

Theoretical studies of such systems have brought interesting
results, some of them being purely quantum, without a classical
counterpart. Among the most beautiful is the binding effect of
curvature due to which infinitely extended regions with hard walls
and a constant width can exhibit localized states. The effect was
in fact observed a long time ago -- see, \eg, \cite{daC1, daC2,
Tol} and references therein -- in formal attempts to justify
quantization on nontrivial manifolds, but only in \cite{ES} it was
placed into a proper quantum-waveguide context and the existence
of geometrically induced discrete spectrum for a curved planar
strip was rigorously proven. This was followed by numerous other
studies in which the results were improved and properties of the
bound states were investigated -- see, \eg, \cite{DE1, GJ, ReBu}.

Much less is known about analogous system in higher dimensions
starting from the physically interesting case of a curved layer.
This may seem strange at a glance, since the leading term of the
effective potential for a general $n$-dimensional manifold in
$\Real^{m+n}$ was computed more than two decades ago \cite{Tol}.
However, going beyond the formal limit of infinitely thin layer
one has to be able to estimate the next terms which is not an easy
task. The aim of the present paper is to stimulate an
investigation of the ``two-in-three" case; we will concentrate at
the simplest case where the deformation of a planar layer is
infinitely smooth and compactly supported.

Let us describe the contents of the paper. In the next section we
will formulate the problem and introduce a technique to handle it
based on a suitable change of coordinates. The main results are
given in Section~3. We will show first that under our assumption
the essential spectrum starts at the first transverse eigenvalue.
Then we will present a variational argument showing that a local
deformation of the layer pushes the bottom of the spectrum below
this value inducing thus a nonempty discrete spectrum. Properties
of these bound states will be discussed elsewhere.

\setcounter{equation}{0}
\section{Formulation of the problem}

Let $\Sigma_0$ be an open set in $\Real^2$; its points will be
denoted by $q=(q^1,q^2)\in\Real^2$. Let a regular and simple
\emph{surface} $\Sigma$ of class $\Diff$ in the space $\Real^3$ be
given by a mapping
\beq p:\Sigma_0\to\Real^3:\{q\mapsto p(q)\in\Sigma\} \eeq
such that the vectors $p_{,\mu}\equiv\partial_{\mu} p:=\partial
p/\partial q^\mu,\: \mu=1,2$, are linearly independent. We have in
mind in this paper surfaces diffeomorphic to the plane, but since
we will use different parametrizations, it is reasonable to
consider $\Sigma_0$ generally as a subset of $\Real^2$. Under the
linear independence condition a unit normal of the surface
\beq n:\Sigma_0\to\Real^3:\left\{q\mapsto n(q):=\frac{p_{,1}\times
p_{,2}}{|p_{,1}\times p_{,2}|} \in\Real^3\right\} \eeq
is a well-defined smooth function, which defines an orientation of
$\Sigma$. Together they determine a \emph{layer} $\Omega$ of a
width $d=2\,a>0$ over the surface $\Sigma$ by virtue of the
mapping
\beq\label{layer} \phi:\Omega_0\to\Real^3:
\{(q,u)\mapsto\phi(q,u):=p(q)+u n(q)\in\Omega\}, \eeq
where $\Omega_0:=\Sigma_0\times (-a,a)$.

\subsection{Properties of the reference surface} \label{surf}

Recall first basic facts about the three fundamental forms of
$\Sigma$. The coefficients of the first fundamental form
$\mathbf{I}$ of the surface can be identified with the covariant
components of its \emph{metric tensor}
\beq
  g_{\mu\nu}:=p_{,\mu}\,.\,p_{,\nu} \qquad g:=\det(g_{\mu\nu})
\eeq
while for the second one, $\mathbf{II}$, we use the common
notation
\beq
  h_{\mu\nu}:=-n_{,\mu}\,.\,p_{,\nu} \qquad h:=\det(h_{\mu\nu}).
\eeq
The Weingarten map $h_\mu^\nu$
(\cf~\cite[Def.~3.3.4~\&~Prop.~3.5.5]{Kli}) determines the
\emph{Gauss curvature} of~$\Sigma$,
$K:=\det(h_\mu^\nu)=\frac{h}{g}$, and its \emph{mean curvature},
$M:=\frac{1}{2}\Tr(h_\mu^\nu)=\frac{1}{2}g^{\mu\nu}h_{\mu\nu}$.
The third fundamental form $\mathbf{III}:=(n_{,\mu}\,.\,n_{,\nu})$
may be expressed by means of the first and second fundamental
forms as follows:


\begin{prop}\label{III}
$\;n_{,\mu}\,.\,n_{,\nu}=-K g_{\mu\nu}+2 M h_{\mu\nu}
=h_{\mu\rho}\,g^{\rho\sigma}\,h_{\sigma\nu}$
\end{prop}
\PF The first relation is equivalent to the identity
$\mathbf{III}-2M\mathbf{II}+K\mathbf{I}=0$ --
\cf~\cite[Prop.~3.5.6]{Kli} or \cite[Problem~7.53]{LR} for a
general dimension. To prove the second one, we use the fact that
$K,M$ are determined by the characteristic equation
\beq\label{CharEq}
  k_\pm^2-2 M k_\pm+K=0,
\eeq
where $k_+,k_-$ are the \emph{principal curvatures}, \ie\/ the
eigenvalues of $h_\mu^\sigma$. Let $T_\sigma^\pm$ be the principal
direction, \ie\/ the eigenvector of $h_\mu^\sigma$, corresponding
to $k_\pm$. Multiplying~(\ref{CharEq}) by this vector we get
$h_\mu^\rho h_\rho^\sigma T_\sigma^\pm-2M h_\mu^\sigma
T_\sigma^\pm +K\delta_\mu^\sigma T_\sigma^\pm=0$. Since the
principal directions forms locally an orthogonal basis in
$\Real^2$, the same must hold in the matrix sense, $h_\mu^\rho
h_\rho^\sigma-2M h_\mu^\sigma+K\delta_\mu^\sigma=0$. The desired
equality is then obtained by multiplication with
$g_{\sigma\nu}$.\qed

\begin{rem} {\rm
We use the standard \emph{summation convention} about repeated
indices; the Greek and Latin ones run through $1,2$ and $1,2,3$,
respectively. The indices are associated with the above
coordinates by $(1,2,3)\leftrightarrow (q^1,q^2,u)$. Furthermore,
upper and lower index denote components of contravariant and
covariant tensors, respectively. Indices are raised and lowered by
the corresponding metric tensor. For instance, the matrix of the
Weingerten map is given by $h_\mu^\nu=h_{\mu\rho}\,g^{\rho\nu}$.
The  same applies to the metric tensor~$G_{\mu\nu}$ in the
layer~$\Omega$ which we shall introduce below.}
\end{rem}
Next we use the Jacobian $g^\frac{1}{2}$ to write down the
invariant surface element
$$
  d\Sigma:=g^\frac{1}{2}d^2q\equiv g^\frac{1}{2}dq^1 dq^2.
$$
It makes it possible define a global quantity characterizing
$\Sigma$, namely the \emph{total curvature} of our surface as
$$
  \Tot(\Sigma):=\int_\Sigma K d\Sigma.
$$
Suppose that $\mathcal{G}\subset\Sigma$ is a region encircled by a
simple closed curve $\mathcal{C}$ of class~$\Diffii$, then the
Gauss-Bonnet theorem \cite[7.6.45]{LR} claims that
\beq\label{GB}
  \Tot(\mathcal{G})+\oint_\mathcal{C}k_g d\ell=2\pi,
\eeq
where $k_g$ is the geodesic curvature of $\mathcal{C}$ (traversed
in the positive sense) and $\ell$ denotes its arc length. For the
purpose of this paper it is important that the geodesic curvature
of a circle in the plane is equal to its reciprocal radius.
Consequently, an infinite surface obtained by a compactly
supported deformation of a plane has $\Tot(\Sigma)=0$.

\subsection{Metric properties of the layer}

It is clear from the definition (\ref{layer}) that the metric
tensor of the layer (as a manifold with a boundary in $\Real^3$)
is of the following form
\beq\label{metricG}
  G_{ij}:=\phi_{,i}\,.\,\phi_{,j}
  \qquad (G_{ij})=
  \left(
  \begin{array}{ccc}
    G_{11} & G_{12} & 0 \\
    G_{12} & G_{22} & 0 \\
    0      & 0      & 1
  \end{array}
  \right),
\eeq
where $G_{\mu\nu}=g_{\mu\nu}-2\,u\,h_{\mu\nu}+u^2\:
 n_{,\mu}\,.\,n_{,\nu}$. In view of Proposition~\ref{III} we can
rewrite the last expression as
\beq\label{matrixG}
  G_{\mu\nu}=g_{\mu\rho}(\delta_\sigma^\rho-u h_\sigma^\rho)
  (\delta_\nu^\sigma-u h_\nu^\sigma),
\eeq
which makes it easy to compute the determinant because the matrix
of the Weingarten map $h_\mu^\nu$ has the principal curvatures
$k_+,k_-$ as eigenvalues. Hence
\beq\label{detG}
  G:=\det(G_{\mu\nu})=g\left[(1-uk_+)(1-uk_-)\right]^2
  =g(1-2Mu+Ku^2)^2,
\eeq
where in the second step we employed the relations $K=k_+ k_-$ and
$M=\frac{1}{2}(k_++k_-)$. As above,
$$
  d\Omega:=G^\frac{1}{2}d\Omega_0\equiv G^\frac{1}{2}d^2q\,du
$$
defines the volume element of the layer.

The ``straightening" transformation employed below requires that
the mapping $\phi$ defining the layer is a diffeomorphism. In view
of the regularity assumptions imposed on $\Sigma$ and the inverse
function theorem it is sufficient that $G_{\mu\nu}$ has an inverse
bounded uniformly in $\Omega$. This imposes a restriction of the
layer thickness $d$. Define $\rho_m:=\left(\max\{
\|k_\nu(q)\|_\infty: \:\nu=\pm,\, q\in\Sigma_0\} \right)^{-1}$. It
follows from~(\ref{matrixG}) that $G_{\mu\nu}$ can be estimated by
the surface metric,
%
\beq\label{C+-} C_- g_{\mu\nu}\leq G_{\mu\nu}\leq C_+ g_{\mu\nu},
\eeq
where the constants $C_\pm:=\left(1\pm a\rho_m^{-1}\right)^2$ are
well defined since $\rho_m>0$ by definition. Hence for a smooth
surface $\Sigma$ the bijectivity of $\phi$ is ensured as long as
$a<\rho_m$.

\subsection{Various expressions of the Hamiltonian}

After these preliminaries let us define the Hamiltonian $\tilde H$
of our model. As we have said, the particle is supposed to be free
within $\Omega$ and the boundary of the layer is a hard wall, \ie,
the wavefunctions should satisfy the Dirichlet boundary condition
there. For the sake of simplicity we set the Planck's constant
$\hbar = 1$ and the effective mass of the electron $m_* =1/2$;
then $\tilde H$ can be identified with the Dirichlet Laplacian
\beq\label{Laplacian}
  \tilde H:=-\Delta_D^\Omega\quad \mathrm{on} \quad\sii(\Omega),
\eeq
which is defined for an open set $\Omega\subset\Real^3$ as the
Friedrichs extension of the operator $-\Delta$ with the domain
$\Comp(\Omega)$ -- \cf~\cite[Sec.~XIII.15]{RS}.

A natural way to investigate the operator (\ref{Laplacian}) is to
pass to the intrinsic coordinates $(q,u)$ in which it acquires the
Laplace-Beltrami form,
\beq\label{hatHamiltonian}
  \hat H:=-G^{-\frac{1}{2}}\partial_i G^{\frac{1}{2}}G^{ij}\partial_j
  \quad \mathrm{on} \quad
  \sii(\Omega_0,G^{\frac{1}{2}}d^2q\,du).
\eeq
To find its action explicitly, we employ~(\ref{metricG}) together
with the expression of the determinant~(\ref{detG}). Then $\hat H$
splits into a sum of two parts,
$$
  \hat H=\hat H_1+\hat H_2,
$$
given by
\beqn
  \hat H_1
  & = &
  -G^{-\frac{1}{2}}\partial_\mu G^{\frac{1}{2}}
  G^{\mu\nu}\partial_\nu
  =-\partial_\mu G^{\mu\nu}\partial_\nu
  -\frac{1}{2}G^{-1}G,_\mu G^{\mu\nu}\partial_\nu \label{hatH1} \\
  \hat H_2
  & = &
  -G^{-\frac{1}{2}}\partial_3 G^{\frac{1}{2}}\partial_3
  =-\partial_3^2-2\,\frac{Ku-M}{1-2Mu+Ku^2}\,\partial_3\, \label{hatH2}.
\eeqn
The above coordinate change is nothing else than the unitary
transformation
$$
  \hat U:\sii(\Omega)\to\sii(\Omega_0,G^{\frac{1}{2}}d^2q\,du):
  \{\psi\mapsto \hat U\psi:=\psi\circ\phi\}
$$
which relates the two operators by $\hat H=\hat U\tilde H \hat
U^{-1}$.

At the same time, it is useful to have an alternative form of the
Hamiltonian which is symmetric w.r.t. $G$ and has the Jacobian
removed from the inner product. It is obtained by another unitary
transformation,
$$
  U:\sii(\Omega)\to\sii(\Omega_0):
  \{\psi\mapsto U\psi:=G^{\frac{1}{4}}\psi\circ\phi\}.
$$
which leads to the unitarily equivalent operator
%
\beq\label{Hamiltonian}
  H:=U\tilde H U^{-1}=
  -G^{-\frac{1}{4}}\partial_\mu G^{\frac{1}{2}}
  G^{\mu\nu}\partial_\nu G^{-\frac{1}{4}}
  -G^{-\frac{1}{4}}\partial_3 G^{\frac{1}{2}}\partial_3 G^{-\frac{1}{4}}
\eeq
with the domain
\beq
  D(H):=\{\psi\in\sobii(\Omega_0)|\ \forall q\in\Sigma_0:
  \psi(q,-a)=\psi(q,a)=0\},,
\eeq
where $\sobii(\Omega_0)$ is the appropriate local Sobolev space in
the sense of \cite[Sec.~XIII.14]{RS}. Commuting $G^{-\frac{1}{4}}$
with the gradient components we cast the
operator~(\ref{Hamiltonian}) into a form which has a simpler
kinetic part but contains an effective potential,
%
\beq\label{Hamiltonian2}
  H = -\partial_\mu G^{\mu\nu}\partial_\nu - \partial_3^2 + V,
\eeq
with
$$ V=F^i_{,i}+F_i F^i\,, \quad F_i:=(\ln G^\frac{1}{4})_{,i}. $$
This expression of the potential is valid for any smooth metric
$G_{ij}$. If we employ the particular form (\ref{metricG}) of the
metric tensor, we can write again (\ref{Hamiltonian2}) as a sum of
two parts, $H\equiv H_1+H_2$ with $V=V_1+V_2$, where
\beqn
  H_1
  & = &
  -\partial_\mu G^{\mu\nu}\partial_\nu+V_1 \label{H1}\\
  V_1
  & := &
  \frac{1}{4}\partial_\mu G^{-1}G^{\mu\nu}G_{,\nu}
  +\frac{1}{16}G^{-2}G{,_\mu} G^{\mu\nu} G_{,\nu} \nonumber \\
  & = &
  -\frac{3}{16}G^{-2}G_{,\mu} G^{\mu\nu} G_{,\nu}
  +\frac{1}{4}G^{-1}G^{\mu\nu}G_{,\mu\nu}
  +\frac{1}{4}G^{-1}G_{,\mu} G^{\mu\nu}_{\ \ ,\nu}  \label{V1} \\
  H_2
  & = &
  -\partial_3^2+V_2 \\
  V_2
  & := &
  \frac{K-M^2}{(1-2Mu+Ku^2)^2} \label{V2}
\eeqn
%

\subsection{Coordinate decoupling}\label{Sec.InfThin}

While the operator $H_1+V_2$ depends on all the three coordinates,
in thin layers its ``leading term" depend on the longitudinal
coordinates $q$ only. The transverse coordinate $u=q_3$ is
isolated in $H_2-V_2= -\partial_3^2$, so up to higher-order terms
in $a$ the Hamiltonian decouples into a sum of the operators
\beqn
  H_q
  & = &
  -g^{-\frac{1}{4}}\partial_\mu g^{\frac{1}{2}}
  g^{\mu\nu}\partial_\nu g^{-\frac{1}{4}}+K-M^2, \\
  H_u
  & = &
  -\partial_3^2.
\eeqn
This observation is behind the formal limit $a\to 0+$ mentioned in
the introduction \cite{daC1, daC2, Tol} in which the transverse
part is thrown away and the thin-layer Hamiltonian is replaced by
the surface operator $H_q$, with the energy appropriately
renormalized. This procedure can be given meaning in the
perturbation-theory framework, in analogy with \cite{DE1}, as we
shall discuss elsewhere.

Here we use it for motivation purposes. The effective ``surface
potential'' $K-M^2$ can be rewritten by means of the principal
curvatures of~$\Sigma$ as follows
\beq\label{K-M^2}
  K-M^2=-\frac{1}{4}(k_+-k_-)^2.
\eeq
If the curvature vanishes at large distances, we get a potential
well which could imply existence of bound states that would
persist in layers of a finite thickness. In distinction to the
``one-in-two" case of a curved planar strip the effective
potential may vanish if the surface is locally spherical,
$k_+=k_-$, however, this cannot happen everywhere at a locally
deformed plane. Let us also remark that similar Laplace-Beltrami
operators penalized by a quadratic function of the curvature lead
on compact surfaces to interesting isoperimetric problems
\cite{Ha, HaLo, EHL}.

In the next section we shall also need the eigenfunctions
$\{\chi_n\}_{n=1}^\infty$ of the transverse operator $H_u$. They
are given by
\beq\label{TransEF}
  \chi_n(u)=\left\{
  \begin{array}{lcl}
    \sqrt{\frac{2}{d}}\cos\kappa_n u
    & \quad\textrm{if}\ & n\ \textrm{is \emph{odd}} \\
    && \\
    \sqrt{\frac{2}{d}}\sin\kappa_n u
    & \quad\textrm{if}\ & n\ \textrm{is \emph{even}}
  \end{array}
  \right.
\eeq
and the corresponding eigenvalues are $\kappa_n^2=(\kappa_1 n)^2$
with $\kappa_1=\frac{\pi}{d}$.

\setcounter{equation}{0}

\section{Spectrum of locally deformed planar layers}

In what follows we shall consider a class of layers over surfaces
which are smooth local deformations of a plane. More specifically,
suppose that the deformed part of the surface is
$\calA\subset\Sigma$ and denote $\supp K \cup \supp M
=p^{-1}(\calA)=:\calA_0$; it is clear that $\Sigma\setminus\calA$
is a plane with a ``hole", not necessarily a simply
connected one.

Let $(X,\delta_{\mu\nu})$ be a natural representation of
$\Sigma\setminus\calA$ in Cartesian coordinates $(x^1,x^2)\in
X\subset\Real^2$ given by an isometry
$$
  \calC: \Sigma_0\setminus\calA_0 \to X:
  \{q\mapsto(x^1,x^2)=:\calC(q)\}.
$$
In view of the compact-support assumption we can choose $r_0>0$ in
such a way that $\calB_{r_0}:=\{w\in\Sigma:|\calC \circ
p^{-1}(w)|\leq r_0\}$ contains $\calA$ and thus
$\Sigma_{r_0}:=\Sigma\setminus\calB_{r_0}$ is the undeformed plane
with the disc of radius $r_0$ removed. It is useful to introduce a
polar-coordinate parametrization of~$\Sigma_{r_0}$ given by the
isometry
%
\beq\label{polar}
  \calP:(r_0,\infty)\times S^1 \to X:
  \{(r,\vartheta)\mapsto (x^1,x^2)=:(r\cos\vartheta,r\sin\vartheta)\};
\eeq
the corresponding metric tensor acquires then the form
$\diag(1,r^2)$.

\subsection{The essential spectrum}

In a planar layer the essential spectrum starts from the lowest
transverse eigenvalue. We will use the standard bracketing
argument  \cite[Sec.~XIII.15]{RS} in combination with the minimax
principle to prove that the same remains true after a compactly
supported deformation.
%
\begin{prop} \label{ess}
$\sigma_\mathrm{ess}(\tilde{H})=[\kappa_1^2,\infty)$.
\end{prop}
\PF We cut the layer $\Omega$ perpendicularly at the boundary of
$\calB_{r_0}$ and impose there the Neumann or Dirichlet condition
respectively; this enables us to squeeze $H$ between a pair of
operators
%
\beq\label{EstEss} H^N_\mathit{int}\oplus H^N_\mathit{ext}\leq H
\leq H^D_\mathit{int}\oplus H^D_\mathit{ext}, \eeq
which have both the form of an orthogonal sum. The spectrum of the
interior parts is purely discrete, so the essential components are
determined by the exterior part only,
$\sigma_\mathrm{ess}(H^\beta_\mathit{int}\oplus
H^\beta_\mathit{ext}) =\sigma_\mathrm{ess}(H^\beta_\mathit{ext})
=\sigma(H^\beta_\mathit{ext})$, $\beta=N,D$. The latter can be
simply localized employing the polar-coordinate
parametrization~(\ref{polar}) of~$\Sigma_{r_0}$. In particular,
let
$$
  U_\calP:\ \sii\left(\Omega_0\setminus p^{-1}(\Sigma_{r_0})
                      \times (-a,a) \right)
  \to \sii\left((r_0,\infty)\times S^1 \times (-a,a)\right)
$$
be the substitution-type unitary operator $(U_\calP\psi)
(r,\vartheta,u):= \psi(\calP^{-1}\circ\calC(q),u)$. It is clear
that the spectrum of the corresponding exterior Hamiltonians
$$
  U_\calP H^\beta_\mathrm{ext} U_\calP^{-1}
  =-\partial_r^2-\frac{1}{r^2}\partial_\vartheta^2
  -\partial_u^2-\frac{1}{4r^2}
$$
contains all points $\kappa_1^2+\epsilon_\beta$, where
$\epsilon_\beta$ belongs to the spectrum of the $s$-wave radial
part, $h^\beta:=-\partial_r^2-(4r^2)^{-1}$ in $\sii(r_0,\infty)$
with the appropriate b.c. at $r=r_0$. We have
$$
  -(\partial_r^2)_N-\frac{1}{4r_0^2}\leq h^N \mathrm{\quad and
  \quad}  h^D \leq-(\partial_r^2)_D
$$
in the sense of quadratic forms and $\inf\sigma_\mathrm{ess}
(-(\partial_r^2)_\beta)= 0$. Since $r_0$ can be chosen arbitrary
large, the claim follows from~(\ref{EstEss}) by the minimax
principle.\qed

\subsection{Existence of Bound States} \label{exbs}

Now comes the main result of this paper. We are going to show that
the conjecture about existence of a discrete spectrum in locally
curved layers formulated in Sec.~\ref{Sec.InfThin} is true, even
for layers which may not be thin. The variational proof of the
following results is based on the idea adapted from \cite{GJ}, see
also~\cite[Thm.~2.1]{DE1}.
%
\begin{thm}\label{ThmExistence}
Suppose that the layer is not planar and the deformation satisfies
the smoothness and compact support assumptions. Then
$\inf\sigma(\tilde H)<\kappa_1^2$.
\end{thm}
\PF
\begin{sloppy}
Denote the norm in $\sii(\Omega_0,G^\frac{1}{2}d^2q\,du)$ as
$\|\cdot\|_G$; then it follows from~(\ref{hatHamiltonian}) that
the quadratic form associated with our Hamiltonian $\hat H$ is
given by
\end{sloppy}
$$
  q[\psi]:=\|\hat H^\frac{1}{2}\psi\|^2_G=q_1[\psi]+q_2[\psi]
$$
where
\beqn
  q_1[\psi]:=\|\hat H_1^\frac{1}{2}\psi\|^2_G
  & = &
  (\psi_{,\mu}, G^{\frac{1}{2}}G^{\mu\nu}\psi_{,\nu}) \\
  q_2[\psi]:=\|\hat H_2^\frac{1}{2}\psi\|^2_G
  & = &
  \|G^{\frac{1}{4}}\psi_{,3}\|^2.
\eeqn
It acts on~$Q(\hat H)$, the quadratic form domain of~$\hat H$. In
order to prove the claim it is sufficient to find a trial function
$\psi\in Q(\hat H)$ such that
$$
  t[\psi]:=q[\psi]-\kappa_1^2\,\|\psi\|^2_G<0.
$$
%

\noindent\textbf{(a)}\,\
We begin the construction of a trial function with
$\psi(q,u):=\varphi(q)\chi_1(u)$, where $\chi_1$ is the lowest
transverse-mode function~(\ref{TransEF}) and $\varphi$ is a
function from the Schwartz space $\mathcal{S}(\Real^2)$, arbitrary
for a moment. It yields
\beqn
  q_1[\psi]
  & = &
  \left(\varphi_{,\mu},\langle G^{\frac{1}{2}}
  G^{\mu\nu}|\chi_1|^2\rangle_u\,\varphi_{,\nu}\right)_q \\
  q_2[\psi]
  & = &
  \left(\varphi,\langle G^{\frac{1}{2}}|\chi'_1|^2\rangle_u
  \,\varphi\right)_q \\
  \|\psi\|^2_G
  & = &
  \left(\varphi,\langle G^{\frac{1}{2}}|\chi_1|^2
  \rangle_u\,\varphi\right)_q
\eeqn
where $\langle\cdot\rangle_u$ means a ``transverse" expectation
and the subscripts $q$, $u$ mark the fact that we integrate w.r.t.
the corresponding coordinate only.

Taking into account the explicit expression (\ref{detG}) for $G$
and using the trivial fact that $|\chi_1|$, $|\chi'_1|$ are
\emph{even} functions and that we integrate over a symmetric
interval $(-a,a)$, and consequently, that we can consider just the
\emph{even} powers of~$u$ in $\langle\cdot\rangle$, we get
$$
  \langle G^{\frac{1}{2}}|\chi'_1|^2\rangle_u
  -\kappa_1^2\,\langle G^{\frac{1}{2}}|\chi_1|^2\rangle_u
  =K g^{\frac{1}{2}};
$$
we have employed at that the identity $\langle u^2(|\chi'_1|^2\!
-\kappa_1^2|\chi_1|^2)\rangle_u =1$. By virtue of~(\ref{C+-}), we
can estimate the remaining term as
\beq\label{q1<} q_1[\psi]\leq C_+
\left(\varphi_{,\mu},g^\frac{1}{2}g^{\mu\nu}
\varphi_{,\nu}\right)_q. \eeq
Suppose now that $\varphi(q)=1$ on $p^{-1}(\calB_{r_0})$ and that
the function is radially symmetric in the sense
$\tilde{\varphi}(r,\vartheta)=\tilde\varphi(r)$, where
$\tilde{\varphi}:=\varphi\circ\calC^{-1}\circ\calP$. Passing then
to the polar coordinates ($g^\frac{1}{2}d^2q=r dr d\vartheta$)
in~(\ref{q1<}) we arrive at
$$
  q_1[\psi]\leq C_+ \int_{\posReal\times S^1}
  |\dot{\tilde{\varphi}}|^2 r\,dr d\vartheta
  =:C_+ \,\|\dot{\tilde{\varphi}}\|_\mathcal{P},
$$
where the dot denotes the derivative w.r.t. $r$. The \rhs of this
inequality depends on the surface geometry through the constant
$C_+$ only. Summing up the results we have
\beq\label{estt}
  t[\psi] \leq C_+ \,\|\dot{\tilde\varphi}\|_\mathcal{P}^2
  +(\varphi,K g^{\frac{1}{2}}\,\varphi)_q\,.
\eeq
%
\noindent\textbf{(b)}\,\
In the next step we shall specify further the function
$\tilde\varphi$ in a way which allows us to make the \rhs
of~(\ref{estt}) arbitrary small. Let us define the family
$\{\tilde\varphi_\sigma:\sigma\in(0,1]\}$ by an external scaling
(in the region $r>r_0$) of a suitable function. The idea is
analogous to ~\cite{GJ} or \cite[Thm.~2.1]{DE1}, however, since we
deal with a two-dimensional integral we have to be more careful
about the decay properties. We can adopt for this purpose the mollifier
employed in~\cite{EV,BCEZ}, which is expressed in terms
of Macdonald functions (or modified Bessel functions in the
terminology of~\cite{AS}) as
$$
  \tilde\varphi_\sigma(r):=\min\left\{1,\frac{K_0(\sigma r)}{K_0(\sigma r_0)}
  \right\}
$$
Since $K_0$ is strictly decreasing, the corresponding
$\psi_\sigma:=\varphi_\sigma\chi_1$ will not be smooth at $r=r_0$
but it remains continuous, hence it is an admissible trial
function as an element of~$Q(\hat H)$.

To estimate the first term at the \rhs\/ of~(\ref{estt}), let us
compute the norm of the scaled function using~\cite[Sec.~9.6]{AS}
and \cite[5.54]{GR}:
\begin{eqnarray*}
  \|\dot{\tilde\varphi}_\sigma\|_\mathcal{P}^2
  & = &
  \frac{2\pi}{K_0(\sigma r_0)^2}\int_{r_0}^\infty\dot K_0(\sigma r)^2\,r\,dr
  =\frac{2\pi}{K_0(\sigma r_0)^2}\int_{\sigma r_0}^\infty K_1(t)^2\,t\,dt \\
  & = &
  \frac{\pi\,(\sigma r_0)^2}{K_0(\sigma r_0)^2}\,
  \left[K_0(\sigma r_0)K_2(\sigma r_0)-K_1(\sigma r_0)^2\right] \\
  & = &
  \frac{\pi\,(\sigma r_0)^2}{K_0(\sigma r_0)^2}\,
  \left[K_0(\sigma r_0)^2+\frac{2}{\sigma r_0}K_0(\sigma r_0)K_1(\sigma r_0)
  -K_1(\sigma r_0)^2\right] \\
  & = &
  -\frac{\pi}{\ln\sigma r_0}\Bigg[-(\sigma r_0)^2 \ln\sigma r_0
  -2\sigma r_0 \ln\sigma r_0 \,\frac{K_1(\sigma r_0)}{K_0(\sigma r_0)}\\
  &&
  \phantom{-\frac{\pi}{\ln\sigma r_0}\Bigg[-(\sigma r_0)^2 \ln\sigma r_0}
  +(\sigma r_0)^2 \ln\sigma r_0 \left(\frac{K_1(\sigma r_0)}
  {K_0(\sigma r_0)}\right)^2\Bigg]
\end{eqnarray*}
Next we use the small-argument asymptotic
expressions~\cite[Sec.~9.6]{AS}{\setlength\arraycolsep{2pt}
\begin{eqnarray*}
  K_0(x) &=& -\ln x+\mathcal{O}(1) \\
  K_1(x) &=& \frac{1}{x}+\mathcal{O}(\ln x)
\end{eqnarray*}
which imply that $x\ln x \,\frac{K_1(x)}{K_0(x)}$ remains bounded
as $x\to 0+$, hence
\beq\label{estphi}
  \|\dot{\tilde\varphi}_\sigma\|_\mathcal{P}^2<\frac{b}{|\ln\sigma r_0|}
\eeq
holds for a positive constant $b$ and $\sigma r_0$ small enough.

\noindent\textbf{(c)}\,\
To handle the second term at the \rhs\/ of~(\ref{estt}) for $
\varphi=\varphi_\sigma$ we employ the dominated convergence
theorem: since $|\varphi_\sigma|\leq 1$ and $\varphi_\sigma\to 1$
pointwise as $\sigma\to 0+$, we have
$$
  (\varphi_\sigma,K g^{\frac{1}{2}}\,\varphi_\sigma)_q\to
  (1,K g^{\frac{1}{2}})_q \equiv \Tot(\Sigma)
$$
by definition. Notice that the total curvature integral is well
defined because the Gauss curvature $K$ of a surface obtained by a
smooth compactly supported deformation of a plane  belongs to
$\si(\Sigma_0,g^\frac{1}{2}d^2q)$. In view of~(\ref{estt}),
(\ref{estphi}) and the last formula we have therefore
\beq\label{estQ1}
  t[\psi_\sigma]\to \Tot(\Sigma)
\eeq
as $\sigma\to 0+$. Recall that in Sec.~\ref{surf} we have used the
Gauss-Bonnet theorem~(\ref{GB}) to show that $\Tot(\Sigma)=0$
holds for surfaces obtained by a smooth local deformation of a
plane. Thus $t[\psi_\sigma]$ can be made arbitrarily small by
choosing $\sigma$ small enough. Since we want to make the form
negative, we have to modify the trial function $\psi_\sigma$
further in analogy with~\cite{GJ}.

\noindent\textbf{(d)}\,\
To this aim we pick $j\in\Comp(\calA_0 \times (-a,a))$ and set
$\Theta:=j^2(\hat H-\kappa_1^2)\psi_\sigma$.
From~(\ref{hatH1})--(\ref{hatH2}) and the fact that the scaling
acts out of the support of the localization function $j$, we
immediately get the following explicit expression
\begin{eqnarray*}
  \Theta(q,u)
  &=&
  j(q,u)^2\,\pi\left(\frac{2}{d}\right)^\frac{3}{2}\!
  \frac{Ku-M}{Ku^2-2Mu+1}\,\sin\kappa_1 u \\
  &=&
  j(q,u)^2\,\pi\left(\frac{2}{d}\right)^\frac{3}{2}\!
  (\ln G^\frac{1}{4})_{,3}\,\sin\kappa_1 u\,.
\end{eqnarray*}
Notice that by construction the function~$\Theta$ does not depend
on~$\sigma$. It is non-zero as an element of $\sii(\Omega_0,
G^\frac{1}{2}d^2q\,du)$ for a non-zero $j$ unless $G$ is
independent of $u$. In view of~(\ref{detG}), however, the last
named situation occurs only if $K,\,M$ are zero identically on the
whole surface which is impossible because $\Sigma$ is not a plane by
assumption.

Since both $\psi_\sigma$ and $\Theta$ belong to~$Q(\hat H)$, we
have
$$
  t[\psi_\sigma+\varepsilon\,\Theta]=t[\psi_\sigma]
  +2\,\varepsilon\,\|j(\hat H-\kappa_1^2)\psi_\sigma\|^2
  +\varepsilon^2\,t[\Theta]\,.
$$
For all sufficiently small \emph{negative} $\varepsilon$ the sum
of the last two terms is negative, and the above arguments shows
that we can choose~$\sigma$ so that $t[\psi_\sigma
+\varepsilon\,\Theta]<0$; recall that the second term on the right
side is independent of~$\sigma$.\qed

\begin{rem}{\rm
Notice that the choice of the Macdonald function $K_0(r)$ for the
mollifier $\tilde\varphi$ in the part \textbf{(b)} is not
indispensable. One can modify this part of the proof, \eg, by
using $e^{-r} \ln r$. However, the choice we made is the most
natural in a sense, because it employs the Green's function kernel
at zero energy.}
\end{rem}

\smallskip

The obtained conclusion about the bottom of the spectrum can be
combined with Proposition~\ref{ess} to get the result announced at
the beginning of Sec.~\ref{exbs}.
\begin{corol}
Let $\Omega$ be a curved layer built over $\Sigma$ which is a
nontrivial, local, and smooth deformation of a plane, with the
half-thickness strictly smaller than the minimum curvature radius
of $\Sigma$ -- \cf(\ref{C+-}). Then $\tilde H$ has at least one
bound state with energy below $\kappa_1^2$.
\end{corol}


\subsection*{Acknowledgment}

The research has been partially supported by GAAS and the Czech
Ministry of Education under the contracts \#1048801 and ME099.



\end{document}